\documentclass[
  aps,
  prl,
  reprint,
  showpacs,
  groupedaddress,
  amsmath,
  amssymb,
  floatfix]{revtex4-1}

\usepackage{graphicx,epsfig,dcolumn,multirow}
\newcolumntype{d}[1]{D{.}{.}{#1}}

\RequirePackage{xspace}

\newcommand{\dis}[1]{\begin{equation}\begin{split}#1\end{split}\end{equation}}
\newcommand{\be}{\begin{equation}}
\newcommand{\ee}{\end{equation}}

\newcommand{\eq}[1]{Eq.~(\ref{#1})}

\def\bea{\begin{eqnarray}}
\def\eea{\end{eqnarray}}

\begin{document}
%\draft

\title{\Large\bf Hidden-sector-assisted 125 GeV Higgs boson
}

\author{ Bumseok Kyae$^{(a)}$\footnote{email: bkyae@pusan.ac.kr}
and Jong-Chul Park$^{(b)}$\footnote{email: jcpark@kias.re.kr} }
\affiliation{$^{(a)}$
Department of Physics, Pusan National University, Busan 609-735, Korea
\\
$^{(b)}$ Korea Institute for Advanced Study, Seoul 130-722, Korea}

%\maketitle

\begin{abstract}

In order to significantly raise the mass of the supersymmetry Higgs boson,
we consider a radiative correction to it by heavy ($\sim 1$ TeV) hidden
sector fields, which communicate with the Higgs through relatively
heavy ``messengers'' (300$-$500 GeV).
The messenger fields ($S$, $\overline{S}$) are coupled to the
Higgs (``$y_HSH_uH_d$,'' $y_H\lesssim 0.7$) and also to hidden
sector fields with a Yukawa coupling of order unity.
The hidden sector fields
are assumed to be large representations of a hidden gauge group,
and so their scalar partner masses can be heavier than other typical
soft scalars in the visible sector.
Even with a relatively small $y_H$ ($\sim 0.2$) or ${\rm
tan}\beta\sim 10$ but without top-stop's considerable
contributions, the radiative correction by such hidden sector fields can
be enhanced enough to yield the $125$ GeV Higgs mass.
\end{abstract}


\maketitle

%%%%%%%%%%%%%%%%%%%%%%%%%%%%%%%%%%%%%%%%%%%%%%%%%%%%%%%%%%%%%%%%%%%%%%%%%%%%%%%%%
%%%%%%%%%%%%%%%%%%%%%%%%%%%%%%%%%%%%%%%%%%%%%%%%%%%%%%%%%%%%%%%%%%%%%%%%%%%%%%%%%

%\section{Introduction}

The solution of the naturalness problem and the gauge coupling
unification are indeed the great achievements of the minimal
supersymmetric standard model (MSSM). By supersymmetry (SUSY) in the MSSM, the smallness of the Higgs mass can be perturbatively
valid up to the fundamental scale, and so the standard model (SM)
can be naturally embedded in a unified theory in the Planck or
string scale. The gauge coupling unification in the MSSM might be
strong evidence for it.

Recently, the ATLAS and CMS collaborations reported the excesses
of events for the $\gamma\gamma$, $ZZ^*\rightarrow 4l$, and
$WW^*\rightarrow 2l2\nu$ channels around 125 GeV invariant mass.
They seemingly imply the presence of the Higgs with 125 GeV mass
at 3$\sigma$ confidence level \cite{ATLAS-CMS}. They might be
accepted as the signals of the MSSM, since the 125 GeV Higgs mass
is still inside the range that the MSSM admits, just assuming the
relatively heavy stop ($\gtrsim$ a few TeV) \cite{MSSM}. Indeed
the radiative correction by the top ($t_{L,R}$) and stop
($\tilde{t}_{L,R}$), and the maximal mixing in $(\tilde{t}_L,
\tilde{t}_R)$ via the SUSY-breaking ``$A$ term'' can raise the Higgs
mass up to 135 GeV in the decoupling limit of the $CP$ odd Higgs
\cite{MSSM}. For 125 GeV Higgs mass, however, the
naturalness of the relatively light Higgs mass in the MSSM would
now become seriously challenged with such a heavy stop.
%
%because the mass bounds for the superparticles of the first two families have already severely rised.

The naturalness problem of the Higgs mass may be much alleviated
in the next-to-minimal supersymmetric standard model (NMSSM)
\cite{nmssm}. In the NMSSM, the ``$\mu$ term'' of the MSSM is
promoted to a trilinear superpotential ``$y_HSH_uH_d$,''
introducing a singlet superfield $S$ \cite{S-ext}. It provides an additional
$F$-term quartic coupling in the Higgs potential apart from the
quartic potential coming from the $D$-term in the MSSM. Thus, the
maximum mass of the lightest Higgs is modified as \dis{
\label{nmssmHiggs} m_h^2\approx M_Z^2{\rm
cos}^22\beta+y_H^2v^2~{\rm sin}^22\beta+{\Delta}m_h^2 ~, } where
$v^2\equiv{v_u^2+v_d^2}\approx (174~{\rm GeV})^2$ and $\Delta
m_h^2$ indicates the radiative correction by the (s)top.
In the MSSM ($y_H=0$), for instance, the needed $\Delta m_h^2|_{\rm top}$
is around $(85~{\rm GeV})^2$ for the $125$ GeV Higgs mass.
To obtain $m_h\approx 125$ GeV with smaller contributions of the
(s)top, $\Delta m_h^2|_{\rm top}\lesssim (85~{\rm GeV})^2$ in the NMSSM;
therefore, the coupling $y_H$ should be larger than 0.5 for ${\rm
tan}\beta >1$. In the NMSSM, however, $y_H$ is stringently
restricted by the Landau pole constraint, $y_H \lesssim 0.7$
\cite{nmssm}. Moreover, only a quite narrow range of ${\rm
tan}\beta$, $1\lesssim {\rm tan}\beta\lesssim 3$, 
which gives almost the {\it maximal} values of ${\rm sin}^22\beta$, is allowed for
${\cal O}(100)$ GeV stop mass.

The fourth generation of the SM particles or extra vectorlike
matter fields, if they exist at the electroweak scale and couple
to the Higgs, can also contribute to $\Delta m_h^2$ by adding
extra radiative corrections to the Higgs mass \cite{extramatt}.
However, the introduction of new colored particles with an order-one
Yukawa coupling to raise the Higgs mass would exceedingly
affect the production rate of two gluons to Higgs ($gg\rightarrow h^0$) as well as the
decay rate of Higgs to two gammas ($h^0\rightarrow\gamma\gamma$): they might result in
immoderate deviations from the SM predictions for their rates,
unless some new invisible Higgs decay channels open
\cite{invisibledecay}. Moreover, for leaving intact the gauge
coupling unification in the MSSM, the newly introduced extra
vectorlike matter fields should compose SU(5) or SO(10)
multiplets.
If the low-energy theory is not embedded in SU(5) [or SO(10)]
grand unifications below the string scale but still keeps the
gauge coupling unification as seen in many string models
\cite{string}, one needs to explore other possibilities for
$m_h\approx 125$ GeV.

In this paper, we will propose a singlet extension of the MSSM to
achieve $m_h\approx 125$ GeV, in which the radiative correction by
hidden sector fields mainly contributes to $\Delta m_h^2$. Since
the radiative correction can be enhanced just with the MSSM
singlets but without the (s)top quark's significant contributions,
the gauge coupling unification is maintained, and also the
naturalness of the SUSY Higgs can still be supported by the
relatively light stop.

%\section{Raising the Higgs Mass}

We consider the following singlet-extended superpotential: \dis{
\label{superPot} W=y_HSH_uH_d + m_SS\overline{S}
+y_N\overline{S}N\overline{N}+m_NN\overline{N} + W_{\rm MSSM}  ,
} where $(S,\overline{S})$ and $(N,\overline{N})$ are the neutral
superfields under the SM gauge symmetry. For the simple
presentation, in \eq{superPot} we do not explicitly write down the
relevant superpotential of the MSSM including the $\mu$ term.
Since the $\mu$ term can actually be generated from the tadpole
term of $S$, the SUSY-breaking $A$ term corresponding to the first
term in \eq{superPot}, the bare $\mu$ term would not be essential.
However, we will keep it to avoid unwanted Pecci-Quinn (PQ)
symmetry breaking at the electroweak energy scale \cite{KNS}. We
suppose that the MSSM singlets $N$, $\overline{N}$ are proper
vectorlike $n$-dimensional representations of a certain hidden
gauge group. They could survive down to low energies by the global
symmetries discussed later. They can communicate with the Higgs
sector through the ``messenger'' fields, $S$ and $\overline{S}$. Note that the messenger in this paper does {\it not} mean the conventional messenger mediating SUSY-breaking from the hidden sector. They just play the role of connecting the Higgs and $(N,\overline{N})$ sectors. 
By redefining the superfields, all the parameters in
\eq{superPot}, i.e. $y_{H}$, $y_N$, and $m_S$ can be made real. We
will discuss later how the effective superpotential \eq{superPot}
can be obtained at low energies.

As in the NMSSM, the superpotential \eq{superPot} yields a quartic
Higgs potential at tree-level, i.e. by $|\partial W/\partial
S|^2$, which could raise the MSSM Higgs mass if the Yukawa
coupling constant $y_H$ was sizable. As mentioned above, however,
the Landau pole constraint restricts the size of $y_H$ to be
smaller than 0.7 \cite{nmssm}. In this paper, we are interested
in the case that $y_H$ is small enough, $0.2\lesssim y_H\lesssim
0.5$, and also $3\lesssim {\rm tan}\beta \lesssim10$.  In this
case, the tree-level quartic contribution ``$y_H^2v^2{\rm
sin}^22\beta$'' in \eq{nmssmHiggs}, which is the dominant
correction in the NMSSM, becomes quite suppressed.
%
%With relatively small $y_H$, thus, the soft mass squared of $S$, $\widetilde{m}_S^2$ would not be severely suppressed than other soft parameters at low energies by the renormalization group (RG) running effect.
%
On the other hand, $y_N$ of order unity can avoid the Landau pole
constraint, because the relatively strong hidden gauge interaction
of $N$, $\overline{N}$ can prevent the blowing-up of $y_N$ at high
energies. We assume that $y_N$ is of order unity at the electroweak
energy scale.

The SUSY mass parameter $m_S$ is assumed to be quite heavier than
the Higgs mass, say, $\gtrsim 300$ GeV. The origin of $m_S$ (and
$m_N$) as well as the ``$\mu$'' in the MSSM can be explained e.g.
by the Giudice-Masiero mechanism \cite{GM}. Since $y_H$ is
relatively small and $m_S$ is quite heavier than the Higgs mass,
the mixing angles between the Higgs and singlet sectors are
expected to be negligible.
If $y_H$ is small enough, $\widetilde{m}_S^2$ does not run with
energy at one-loop level.

As mentioned above, $y_N$ is of order unity. Accordingly, the
soft mass squared of $\overline{S}^2$,
$\widetilde{m}_{\overline{S}}^2$ can be suppressed or even become
negative at low energies by the renormalization group (RG) running
effect. We will assume $|\widetilde{m}_{\overline{S}}|^2\lesssim 
m_S^2$. On the other hand, the soft masses of $N$ and
$\overline{N}$, $\widetilde{m}_{N}$ and
$\widetilde{m}_{\overline{N}}$ can be quite heavier than other
soft masses including $\widetilde{m}_{S}$ at low energies.
Actually, it is also possible by the RG effect for
$\widetilde{m}_{N}$, $\widetilde{m}_{\overline{N}}$, since the
superfields $N$ and $\overline{N}$ are nontrivial
representations under a non-Abelian gauge group in a hidden
sector. We will set
$\widetilde{m}_{N}=\widetilde{m}_{\overline{N}}$, which would be
reasonable in large classes of supergravity (SUGRA) models, if there
is no other Yukawa interaction.
Just for the simplicity of our future calculations, but considering
the above discussions, we suppose the following hierarchy among
the mass parameters in this model: \dis{ \label{hierarchy}
\widetilde{m}_{\overline{S}}^2 \lesssim  \mu^2\lesssim m_{3/2}^2, 
m_S^2 \lesssim \widetilde{m}_{S}^2 \lesssim \widetilde{m}_N^2 , m_N^2 ,}
where $m_{3/2}^2$ collectively denotes other typical soft
parameters. 
Note that even with
this hierarchy, the scalar component of $\overline{S}$, i.e.,
$\widetilde{\overline{S}}$ is still much heavier than the Higgs
because of the SUSY mass $m_S$. Since $m_S$ and $m_N$  both are much heavier than the Higgs mass, there is no ``singlet-ino'' (fermionic component of a singlet superfield) lighter than the Higgs. 

The relevant scalar potential is given by
\begin{eqnarray}
&&V_{\rm F}=\left| y_HH_uH_d+m_S\widetilde{\overline{S}}\right|^2+\left|m_S\widetilde{S} \right|^2
\nonumber \\
&&\qquad~ +\left|y_H\widetilde{S}\right|^2\left(|H_u|^2+|H_d|^2\right) , 
\\
&&V_{\rm soft}=\widetilde{m}_S^2|\widetilde{S}|^2+
\widetilde{m}_{\overline{S}}^2|\widetilde{\overline{S}}|^2
+m_1^2|H_d|^2+m_2^2|H_u|^2
\nonumber \\
&&\quad\quad 
+\left(y_Hm_{3/2}^A\widetilde{S}H_uH_d
+m_Sm_{3/2}^B\widetilde{S}\widetilde{\overline{S}} +{\rm h.c.}\right) , \nonumber
\end{eqnarray}
where $m_{3/2}^A$ and $m_{3/2}^B$ indicate the soft parameters of
$A$ and $B$ terms, respectively. Here we ignored the $\mu$
term due to its relative smallness, and set $\langle
\widetilde{N}\rangle=\langle\widetilde{\overline{N}}\rangle=0$ due
to their heavy masses.
With the hierarchy in \eq{hierarchy}, the equations of motion for $\widetilde{S}$, $\widetilde{\overline{S}}$ yield
\dis{ \label{minimum}
&\langle\widetilde{S}\rangle\approx -\frac{y_H(m^A_{3/2}-m^B_{3/2})}{\widetilde{m}_S^2}~\langle H_uH_d\rangle ,
\\
&\qquad~~~ \langle\widetilde{\overline{S}}\rangle\approx
-\frac{y_H}{m_S}\langle H_uH_d\rangle . } 
%
%For small $y_H$ and heavy
%$\widetilde{m}_S$, thus, the correction to the $\mu$ becomes small.

Since the superfields $N$, $\overline{N}$ couple to $\overline{S}$
in \eq{superPot} and its vacuum expectation value (VEV) should be
related to the Higgs VEVs at the minima of $(\widetilde{S},
\widetilde{\overline{S}})$, $N$ and $\overline{N}$ can get
additional SUSY masses once the Higgs fields develop VEVs. The
relevant diagrams are displayed in FIG. 1, where $N$,
$\overline{N}$ couple to the Higgs $H_u$, $H_d$ through the 
off-shell $\widetilde{\overline{S}}$. Thus, the mass squared of
$(N,\overline{N})$ becomes \dis{ \label{M_N} &~M_N^2=\left[
m_N-y_Hy_N\frac{\langle H_uH_d\rangle}{m_S}\right]^2
\\
&\approx m_N^2-2\left(y_Hy_N\frac{m_N}{m_S}\right)\langle H_uH_d\rangle ~. }
%Since $m_S$ is much heavier than the Higgs mass, the mass term $(y_Hy_N\langle H_uH_d\rangle/m_S)N\overline{N}$ can be a local operator.
%
%where the VEVs of Higgs, $\langle H_{u,d}\rangle$ are taken to be in the real directions.
%
%
By rolling up the $N$, $\overline{N}$ lines of FIG. 1-(a), the
radiative Higgs potential or mass can be generated as seen in FIG.
1-(b).
%
%
%%%%%%%%%%%%%%%%%%%%%%%%%%%%%%%%%%%%%%%%%%%%%%%%%%%%%%%%%%%%
\begin{figure}
\begin{center}
\includegraphics[width=0.8\linewidth]{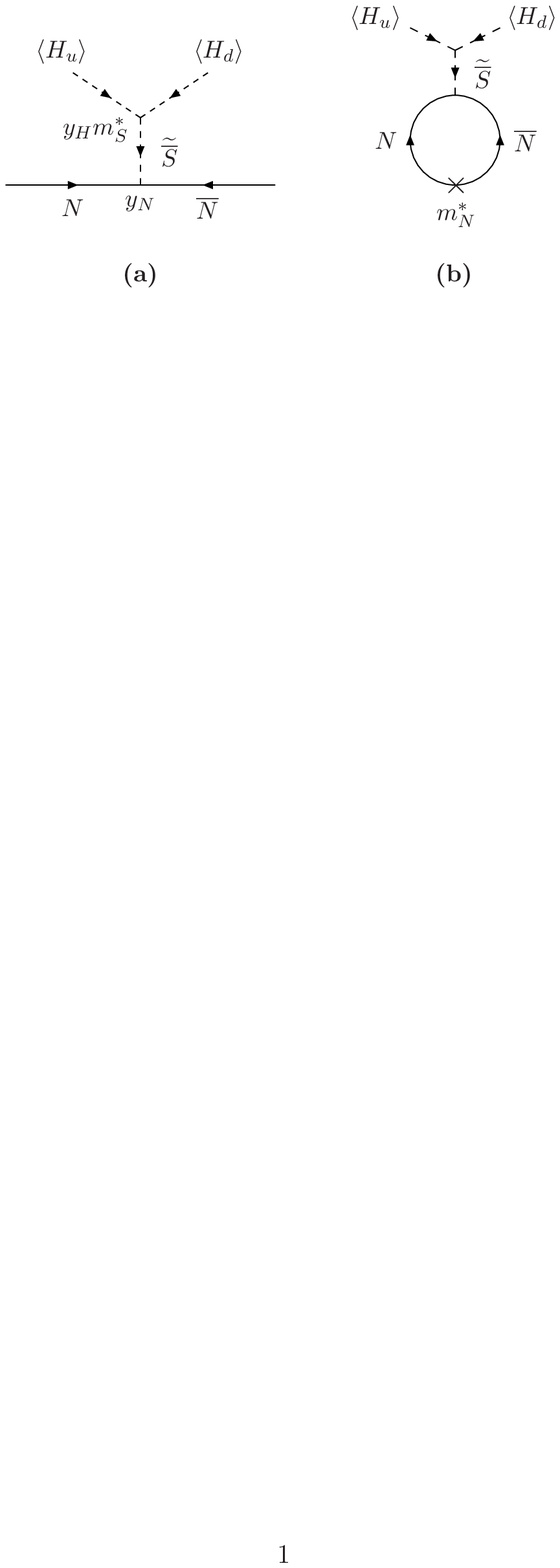}
\end{center}
\caption{{\bf (a)} Additional SUSY mass of the superfields $N$,
$\overline{N}$ generated by the Higgs VEVs. {\bf (b)} A
contribution to the radiatively induced effective Higgs potential.
The trilinear scalar couplings in {\bf (a)} and {\bf (b)} come
from the cross term of $|\partial{W}/\partial{S}|^2$.
}
\label{Fig1}
\end{figure}
%%%%%%%%%%%%%%%%%%%%%%%%%%%%%%%%%%%%%%%%%%%%%%%%%%%%%%%%%%%
%%%%%%%%%%%%%%%%%%%%%%%%%%%%%%%%%%%%%%%%%%%%%%%%%%%%%%%

\begin{figure}
\begin{center}
\includegraphics[width=0.85\linewidth]{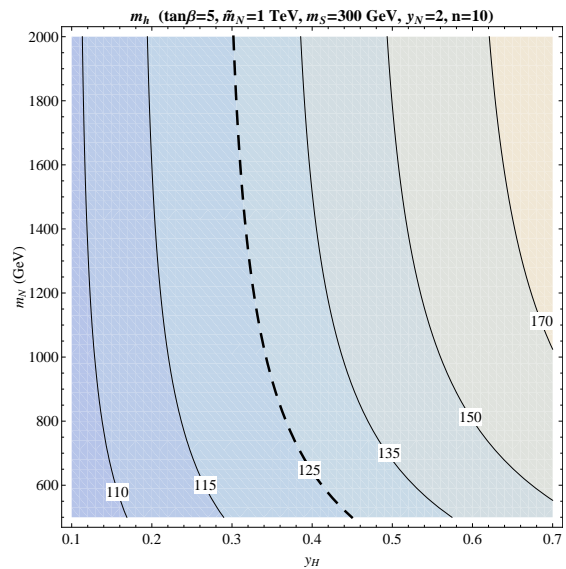}
\end{center}
\caption{Contour plots for the lightest Higgs mass $m_h$ in the
$y_H - m_N$ plane. We set $\Delta m_h|_{\rm top}^2=(66~{\rm GeV})^2$, which corresponds to $\widetilde{m}_t\approx 500$ GeV at two-loop level, but turn off the mixing effect. The tree-level contribution from the NMSSM is ignored. 
We fix the other parameters as shown in the figure. The thick dashed
line corresponds to $m_h = 125$ GeV.}
\label{Fig_mN1}
\end{figure}

\begin{figure}
\begin{center}
\includegraphics[width=1\linewidth]{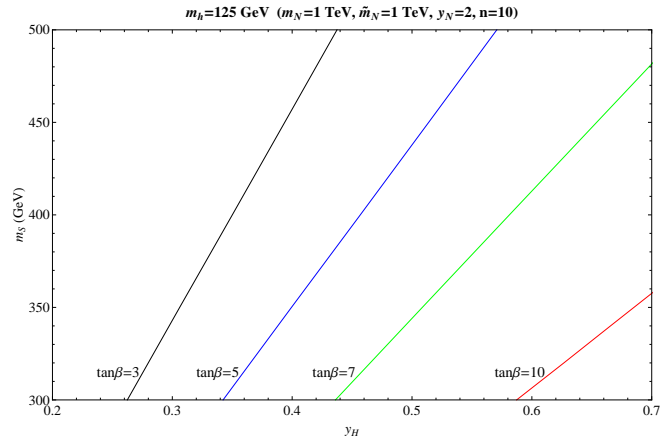}
\end{center}
\caption{Lightest Higgs mass $m_h = 125$ GeV lines for various
$\tan \beta$s in the $y_H - m_S$ plane. 
We set $\Delta m_h|_{\rm top}^2=(66~{\rm GeV})^2$, which corresponds to $\widetilde{m}_t\approx 500$ GeV at two-loop level, but turn off the mixing effect. The tree-level contribution from the NMSSM is ignored. 
The other parameters are fixed as shown in the figure.}
\label{Fig_TanBeta1}
\end{figure}

The scalar components of $N$, $\overline{N}$ possess, of course,
the additional mass terms by SUSY-breaking effects.
%
%As mentioned above, we assume that such soft masses, $\widetilde{m}_{N}$ and  $\widetilde{m}_{\overline{N}}$
%are quite larger than other soft masses at low energies.
%
Due to the mass difference between the fermionic and the bosonic
modes of $N$, $\overline{N}$, the one-loop effective Higgs
potential is generated after integrating out the quantum
fluctuations of $N$, $\overline{N}$ \cite{CW}: 
\begin{eqnarray} \label{1-loopPot} 
\Delta V_{\rm
1-loop}=&&\frac{n}{16\pi^2}\bigg[\left(M_N^2+\widetilde{m}_{N}^2\right)^2\left\{{\rm
log}\left(\frac{M_N^2+\widetilde{m}_{N}^2}{\Lambda^2}\right)-\frac32\right\}
\nonumber \\
&&\quad -M_N^4\left\{{\rm
log}\left(\frac{M_N^2}{\Lambda^2}\right)-\frac32\right\} \bigg] ,
\end{eqnarray} 
where $\Lambda$ denotes a renormalization mass scale. {\it This
effective Higgs potential by the hidden sector fields is valid
only in the energy scales below the messenger scale $m_S$}: above $m_S$ scales the
nonrenormalizable operator suppressed with $m_S$ in \eq{M_N}
cannot be regarded as a local operator any longer. With
\eq{1-loopPot}, the radiative mass correction can be estimated as
\dis{ \label{loopCorr} 
\Delta m_h^2
%=\frac{n}{2\pi^2}\frac{M_N^2}{v^2}\left(M_N-m_N\right)^2{\rm log}\frac{M_N^2+\widetilde{m}_N^2}{M_N^2}
%\\
%&\approx
=\frac{n}{4\pi^2}\left(y_N\frac{m_N}{m_S}\right)^2\left(y_H^2v^2{\rm sin}^22\beta\right) {\rm log}\frac{m_N^2+\widetilde{m}_N^2}{m_N^2} ~.
}
Compared to the case by the (s)top, which is the
dominant correction in the MSSM, 
\dis{ \Delta
m_{h}^2|_{\rm top}=\frac{3}{4\pi^2}(y_tm_t)^2{\rm sin}^2\beta~{\rm
log}\frac{m_t^2+\widetilde{m}_t^2}{m_t^2} ~, } where $y_t$ stands
for the top quark Yukawa coupling, $m_t=y_tv_u$
%and $\widetilde{m}_t$ indicate the top and stop mass, respectively. We
% and we ignored
and the mixing effect by the $A$ term is ignored, we have several
advantages to enhance the radiative correction to the Higgs mass in \eq{loopCorr}:
we can take larger $n$ ($>3$) and heavier $\widetilde{m}_N^2$
($>\widetilde{m}_t^2$), leaving intact the naturalness issue
associated with the stop mass. Moreover, we can quite efficiently
raise the Higgs mass using the factor $(m_N/m_S)^2$ in
\eq{loopCorr}. As a result, the Higgs mass can reach 125 GeV for 
$0.2\lesssim y_H\lesssim 0.7$ and $3\lesssim {\rm tan}\beta\lesssim 10$ with the relatively light stop ($\widetilde{m}_t\approx 500$ GeV). See FIGs. \ref{Fig_mN1} and \ref{Fig_TanBeta1}.

%
%%%%%%%%%%%%%%%%%%%%%%%%%%%%%%%%%%%%%%%%%%%%%%%%%%%%%%%%%%%%
%%%%%%%%%%%%%%%%%%%%%%%%%%%%%%%%%%%%%%%%%%%%%%%%%%%%%%%%%%%%

%%%%%%%%%%%%%%%%%%%%%%%%%%%%%%%%%%%%%%%%%%%%%%%%%%%%%%%%%%%%
%%%%%%%%%%%%%%%%%%%%%%%%%%%%%%%%%%%%%%%%%%%%%%%%%%%%%%%%%%%%

In fact, large mixing in $(\tilde{t}_L, \tilde{t}_R)$ through the $A$ term, 
$X_t/m_{\tilde{t}}\equiv (A_t-\mu{\rm cot}\beta)/m_{\tilde{t}}\approx 2$, where $m_{\tilde{t}}\equiv\sqrt{m_t^2+\widetilde{m}_t^2}$, is very helpful for raising the Higgs mass with relatively light stop.
Particularly, the maximal mixing ($X_t/m_{\tilde{t}}\approx\sqrt{6}$) can push the Higgs mass up to 135 GeV without any other help \cite{MSSM}. 
As the mixing deviates from the maximal mixing, however, the enhancement effect drops rapidly. Without the mixing effect, thus, there is no way to raise the Higgs mass except for increasing the stop mass:
the stop mass needed to achieve 125 GeV Higgs is above a few TeV \cite{MSSM}. However, such a heavy stop mass gives rise to fine-tuning of $0.1$ or $0.01$ percent level among the soft parameters in order to get the Z boson mass of 91 GeV.    
In our analysis, we turn off the mixing effect and set $\Delta m_h^2|_{\rm top}$ by the (s)top to be around $(66~{\rm GeV})^2$, which corresponds to  $\widetilde{m}_t\approx 500$ GeV when the two-loop corrections also included.    

In the parameter ranges we have discussed so far, $\Delta m_h^2$ in \eq{loopCorr} dominates over the tree-level addition, $y_H^2 v^2{\rm
sin}^22\beta$. Normally a one-loop correction is hard to win a tree-level result without extending a system. In our case, it is possible by introducing the $(N,\overline{N})$ sector, which does not participate in the tree-level addition. In our analysis, we do not  include the tree-level addition for increasing the Higgs mass. 

Apart from \eq{loopCorr} we have another part containing ${\rm log}\Lambda$ in the second derivative of the potential \eq{1-loopPot}, 
which renormalizes a soft parameter as in the radiative correction by the (s)top. In this case it does the ``B$\mu$'' parameter ($\equiv m_3^2$). [If one integrates out $(N,\overline{N})$ first, this part also contributes to \eq{minimum} through the induced tadpole potential for ${\widetilde{\overline{S}}}$.
Inserting the extremum conditions for $\widetilde{S}$, $\widetilde{\overline{S}}$ in the potential, however,  gives the same result discussed here.]
Together with the RG solution of the $m_3^2$ appearing in the tree-level potential, it makes a correction to $m_3^2$: 
\dis{
\delta m_3^2\approx &\frac{n}{8\pi^2}y_Hy_N\frac{m_N}{m_S}\bigg[m_N^2~{\rm log}\left(\frac{m_N^2+\widetilde{m}_N^2}{m_N^2}\right) \\
& +\widetilde{m}_N^2\left\{{\rm log}\left(\frac{m_N^2+\widetilde{m}_N^2}{m_S^2}\right)-1\right\}\bigg] .
}
%where we set the cut-off as the messenger scale $m_S$.
%
The parameters considered so far let  $\delta m_3^2$ inside the range of (600 GeV)$^2$ -- (1 TeV)$^2$.

%\section{The Model}

The effective superpotential \eq{superPot} can be derived, e.g., by
employing the following 
superpotential: 
\dis{ \label{UV} 
&\qquad\quad W_{\rm UV}=y_HSH_uH_d+y_N\overline{S}N\overline{N}
\\
&+\frac{f_1}{M_P}\Sigma_1^2H_uH_d+\frac{f_2}{M_P}\Sigma_2^2N\overline{N}+\frac{f_3}{M_P}\Sigma_3^2S\overline{S}
\\
&+\frac{g_1}{M_P}\Sigma_3\Sigma_1\overline{\Sigma}_1^2
+\frac{g_2}{M_P}\Sigma_3\Sigma_2\overline{\Sigma}_2^2+\frac{g_3}{M_P}\Sigma_3^2\overline{\Sigma}_3^2
}
where $y_{H}$, $y_N$, $f_i$, and $g_i$ ($i=1,2,3$) are dimensionless couplings, and $M_P$ denotes the
reduced Planck mass ($=2.4\times 10^{18}$ GeV). The form of the
 superpotential in \eq{UV} can be
controlled by the two continuous global symmetries, U(1)$_{\rm R}$
and U(1)$_{\rm PQ}$. 
The U(1)$_{\rm R}$ and U(1)$_{\rm PQ}$ charge assignments for the superfields appearing in the superpotential \eq{UV} are
presented in TABLE I.

\begin{table}[!h]
\begin{center}
\begin{tabular}
{c|cccccc|cccccc} 
{\rm Superfields}  &   $H_u$   &
 $H_d$  &  ~$N$  & ~$\overline{N}$ & ~$S$  &
 $\overline{S}$  & ~$\Sigma_1$  & ~$\Sigma_2$ & ~$\Sigma_3$ & ~$\overline{\Sigma}_1$ & ~$\overline{\Sigma}_2$ & ~$\overline{\Sigma}_3$   \\
\hline
U(1)$_{\rm R}$ & ~$0$ & ~$0$ & ~$0$ & ~$0$
 & ~$2$ & ~$2$ & ~$1$  & ~$1$ & $-1$& ~$1$& ~$1$& ~$2$ \\
U(1)$_{\rm PQ}$ & $-\frac14$ & $-\frac14$ & ~$\frac12$ & ~$\frac12$ & ~$\frac12$ & $-1$ & ~$\frac14$ & $-\frac12$ & ~$\frac14$& $-\frac14$& ~$\frac18$& $-\frac14$
\end{tabular}
\end{center}\caption{R and Pecci-Quinn charges of the superfields. The MSSM
{\it matter} superfields carry the unit R charges, and  also the PQ charges of $1/8$. $N$ and $\overline{N}$ are assumed to be proper
$n$-dimensional vectorlike representations of a hidden gauge
group, under which all the MSSM fields are neutral.
$\Sigma$s and $\overline{\Sigma}$s carry some $Z_2$ charges.
}\label{tab:Qnumb}
\end{table}

The $A$ terms corresponding to the $g_{1,2,3}$ terms in \eq{UV}, the soft mass terms, etc. in the Lagrangian admit the VEVs of $\Sigma_{1,2,3}$ and $\overline{\Sigma}_{1,2,3}$ of order $\sqrt{m_{3/2}M_P}$ ($\sim 10^{10}$ GeV) \cite{422}, assuming the gravity-mediated SUSY-breaking scenario. 
From the $f_{1,2,3}$ terms in \eq{UV}, thus, $\mu$ in the MSSM, and $m_N$ and $m_S$ in \eq{superPot} are generated, which are of order $m_{3/2}$. Due to the VEVs of $\Sigma$s and $\overline{\Sigma}$s, the U(1)$_{\rm R}$ and U(1)$_{\rm PQ}$ are completely broken at the intermediate scale.  

$\Sigma$s and $\overline{\Sigma}$s in \eq{UV} carry some accidental $Z_2$ charges. As a result, the domain wall problem would arise, if the $Z_2$s are broken after inflation. We assume that such discrete symmetries are already broken before or during inflation. 
If the reheating temperature is lower than $10^9$ GeV, the $Z_2$-breaking vacuum still remains the minimum of the potential also after inflation over \cite{422}.

%\section{Conclusion}

In conclusion, 
we have discussed the possibility that the SUSY Higgs mass increases through the radiative correction by 1 TeV scale hidden sector fields, which can communicate with the Higgs via 
the messenger fields with around 300$-$500 GeV masses. We pointed
out that even for $0.2\lesssim y_H\lesssim 0.5$ or $3\lesssim {\rm
tan}\beta\lesssim 10$, which is the excluded region in the NMSSM, 125 GeV Higgs mass can be naturally explained
in a broad parameter space with relatively light stop masses ($\approx 500$ GeV) but without their mixing effect. 
The fine-tuning problem associated with the light Higgs and heavy soft scalars 
can be avoided, basically because the mass parameters considered for 125 GeV Higgs are all just around a few hundred GeV to 1 TeV.
Consequently, the naturalness and gauge coupling
unification in SUSY models can still be quite valid.

\acknowledgments

This research is supported by Basic
Science Research Program through the National Research Foundation
of Korea (NRF) funded by the Ministry of Education, Science and
Technology (Grant No. 2010-0009021), and also by Korea Institute
for Advanced Study (KIAS) grant funded by the Korean government
(MEST).

%%%%%%%%%%%%%%%%%%%%%%%%%%%%%%%%%%%%%%%%%%%%%%%%%%%%%%%%
%%%%%%%%%%%%%%%%%%%%%%%%%%%%%%%%%%%%%%%%%%%%%%%%%%%%%%%%%

\end{document}